\documentclass[12pt]{article}

\usepackage{latexsym}
\usepackage{graphicx}
\usepackage{amsmath}
\usepackage{amsfonts}
\usepackage{amssymb}
\numberwithin{equation}{section}

\newcommand\bea{\begin{eqnarray}}
\newcommand\eea{\end{eqnarray}}
\newcommand\beq{\begin{equation}}
\newcommand\eeq{\end{equation}}

\newcommand{\bib}{\bibitem}
\newcommand{\email}[1]{Electronic mail: \tt #1}
\newcommand{\mT}{\mathcal{T}}
\def\nn{\nonumber}

\def\f{\frac}

\def\d{\delta}

\def\g{\gamma}

\def\om{\omega}

\def\cT{\mathcal{T}}
\def\bl{{\bf{l}}}

\begin{document}
\newcommand{\emailroy}{\email{dibyendu@rri.res.in}}
\newcommand{\emaildabhi}{\email{dabhi@rri.res.in}}
\newcommand{\add}{Raman Research Institute, Bangalore 560080, India.}
\title{Heat transport in ordered harmonic lattices}
\author{Dibyendu Roy \thanks{\add\ \emailroy} and Abhishek Dhar \thanks{\add\ \emaildabhi}} 

\maketitle
\begin{abstract}
We consider heat conduction across an ordered oscillator chain with harmonic
interparticle interactions and also onsite harmonic potentials. 
The onsite spring constant is the same for all sites excepting the
boundary sites. The chain is connected to Ohmic heat reservoirs at
different temperatures. We use an approach following from a direct
solution of the Langevin equations of motion. This works both in the
classical and quantum regimes. In the classical case we obtain an exact
formula for the heat current in the limit of system size $N \to
\infty$. In special cases this reduces to
earlier results obtained by Rieder, Lebowitz and Lieb and by Nakazawa.
We also obtain results for the quantum mechanical case where we study the
temperature dependence of the heat current.  
We briefly discuss results in higher dimensions.
\end{abstract}
\vskip .5 true cm


\newpage

\section{Introduction}
\label{sec:Intro}

Rieder, Lebowitz and Lieb \cite{Rieder67} (RLL) considered the problem of heat
conduction 
across a one-dimensional ordered harmonic chain connected to
stochastic heat baths at the two ends. The main results of this paper
were: (i) the temperature in the bulk of the system was a constant
equal to the mean of the two bath temperatures, (ii) the heat
current approaches a constant value for large system sizes and an
exact expression for this was obtained. RLL considered the case where
only interparticle potentials were present. Nakazawa \cite{Nakazawa} (N)
extended these results to the case with a constant onsite harmonic
potential at all 
sites and also to higher dimensions. 

The approach followed in both the RLL and N papers was to obtain the
exact nonequilibrium stationary  state measure which, for this
quadratic problem, is a Gaussian distribution. A complete solution for
the correlation matrix was obtained and from this one could obtain both
the steady state temperature profile and the heat current.  

In this paper we use a different formalism to calculate the heat
current in ordered harmonic lattices connected to Ohmic reservoirs
(for a classical system this is white noise Langevin dynamics). The
formalism has been discussed 
in detail in \cite{dhar01,DharRoy06} and follows from a direct solution, by
Fourier transforms, of the
Langevin equations of motion. A general formal expression for the heat
current can be 
obtained and this has the form of an integral of the heat transmitted
over all frequencies. 
For the one-dimensional case this expression for current was first
obtained by Casher and Lebowitz \cite{Casher71} using different methods.   
An advantage of the approach used here is that it  can be easily generalized to
the quantum mechanical regime
\cite{DharRoy06,zurcher90,saito00,dhar03,segal03}. 
As shown in \cite{DharRoy06} the results obtained are identical to 
those obtained using the nonequilibrium Green's function method
\cite{yamamoto06,wang06}.  
Here we show how exact expressions for the asymptotic current ($N\to
\infty$) can be 
obtained  from this approach. We also briefly discuss the model in the quantum
regime and extensions to higher dimensions. 

The model we consider here is  slightly different 
from the Nakazawa model. We consider the pinning potentials at the
boundary sites to be different from the bulk sites. This allows us to
obtain both the RLL and N results as limiting cases. Also it seems
that this model more closely mimics the experimental situation. In experiments
the boundary sites would be interacting  
with fixed reservoirs which can be modeled by an effective spring
constant that is expected to be different from the interparticle
spring constant in the bulk. We also note here that the constant
onsite potential 
present along the wire relates to experimental situations such as that
of heat transport in a molecular wire attached to a substrate  or, in the
two-dimensional case, a monolayer on a substrate. Another example
would be the heat current contribution from the optical modes of a
polar crystal.

The paper is organized as follows. In Sec.\ref{sec:Class}
we introduce the model and derive the main results of the paper.
In Sec.\ref{sec:Quantum} we present results for the quantum
mechanical case and in Sec.\ref{sec:Ddim} generalizations to the higher
dimensional case. Finally we conclude with a short
discussion in Sec.\ref{sec:concl}. 

\section{Model and results in the classical case}
\label{sec:Class}
\begin{figure}[htb]
\vspace{0cm}
\includegraphics[width=14cm]{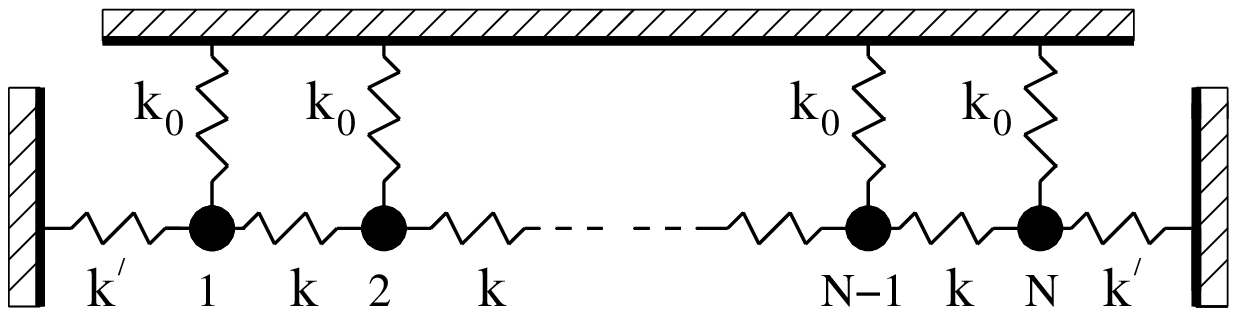}
\caption{A schematic description of the model.}
\label{fig1}
\end{figure}
We consider $N$ particles
of equal masses $m$  connected to each other by harmonic springs of
equal spring constants $k$.  The
particles are also pinned by onsite  quadratic potentials with 
strengths $k_o$ at all sites except the boundary sites where the
pinning strengths are $k_o+k'$ [see Fig.~(\ref{fig1})].  
The Hamiltonian of the harmonic chain is thus:
\begin{eqnarray}
H=\sum_{l=1}^N [\f{1}{2}m \dot{x}_l^2 +\f{1}{2}k_o{x}_l^2]+\sum_{l=1}^{N-1}\f{1}{2}k(x_{l+1}-x_l)^2+ \f{1}{2}k'(x_1^2+x_N^2) ~,\label{ham}
\end{eqnarray}
where $x_l$ denotes the displacement of the particle at site $l$ from
its equilibrium position. The particles $1$ and $N$ at the two ends are
immersed in heat baths at temperature $T_L$ and $T_R$
respectively. The heat baths are assumed to be modeled by Langevin
equations corresponding to Ohmic baths. In the classical case the 
steady state heat current from left to right reservoir is given by
\cite{dhar01,Casher71} :
\bea
J_{C}&=& \f{k_B(T_L-T_R)}{\pi} \int_{-\infty}^\infty d \omega~ \cT_N (\omega)
  ,~~ \label{JSCR} \\
{\rm{where}}~ \cT_N(\omega) &=& \gamma^2 \omega^2 |G_{1N}|^2~, \nn \\    
G&=& [-m\omega^2 I +\Phi - \Sigma]^{-1}~, \nn \\
\Phi_{lm}&=&(k+k'+k_o)~\d_{l,m}-k~\d_{l,m-1}~~~~~~~{\rm{for}}~~~l=1~,\nn\\ 
&=&-k~\d_{l,m+1}+(2k+k_o)~\d_{l,m}-k~\d_{l,m-1}~~~~{\rm{for}}~~~2\leq
l \leq N-1~,\nn\\
&=&(k+k'+k_o)~\d_{l,m}-k~\d_{l,m+1}~~~~~~~{\rm{for}}~~~l=N~,\nn \\
\Sigma_{lm}&=& i \gamma \omega \delta_{lm} [\delta_{l1}+ 
  \delta_{lN}]~, \nn
\eea 
and $I$ is a unit matrix.
We now write $G=Z^{-1}/k$, where $Z$
is a tri-diagonal matrix with 
$Z_{11}=Z_{NN}=(k+k_o+k'-m\om^2-i\g \om)/k$, all other diagonal
elements equal to~ $2+k_o/k-m \om^2/k$ and all off-diagonal elements
equal to $-1$. 
Then it can be shown easily that $|G_{1N}(\om)|=1/(k~|\Delta_N|)$ where
$\Delta_N$ is the determinant of the matrix $Z$. This is
straightforward to obtain and after some rearrangements we get:
\bea
\Delta_N&=&[a(q)\sin Nq +b(q) \cos Nq]/\sin q~, \\
{\rm where}~~ a(q)&=& [2-\f{\g^2 \om^2}{k^2}+\f{{k'}^2}{k^2}-\f{2k'}{k}]\cos q +\f{2 k'}{k}-2-\f{2 i \g \om}{k}[1+(\f{k'}{k}-1)\cos q]~,\nn\\ 
 b(q)&=&[\f{\g^2
    \om^2}{k^2}-\f{{k'}^2}{k^2}+\f{2k'}{k}]\sin q +\f{2 i \g
  \om}{k}(\f{k'}{k}-1)\sin q~,\nn 
\eea
and $q$ is given by the relation $2k\cos q = -m {\om}^2 +k_o+2k$. 
This relation implies that for frequencies outside the phonon band 
$k_o \leq m \om^2 \leq k_o+2k$ the wavevector $q$ becomes imaginary
and hence the transmission coefficient $\cT (\omega)$ decays
exponentially with $N$. Hence for large $N$ we need only consider the
range $0<q<\pi$ and the current is given by:
\bea
J_{C}&=& \f{2\gamma^2 k_B (T_L-T_R)}{k^2\pi} \int_{0}^\pi dq |\f{d \omega}{dq}| ~ \f{\omega^2_q}{|\Delta_N|^2}~, \label{JSCR2} 
\eea
with $m \omega_q^2=k_o+2 k [1-\cos{(q)}]$.
Now we state the following result:
\bea 
\lim_{N \rightarrow \infty}\int_{0}^{\pi} dq \f{g_1(q)}{1+g_2(q)\sin
  Nq}=\int_{0}^{\pi} dq \f{g_1(q)}{[1-g^2_2(q)]^{1/2}}~,\label{iden} 
\eea
where $g_1(q)$ and $g_2(q)$ are any two well-behaved functions. 
This  result can be proved by making an
expansion of the factor $1/[1+g_2(q) \sin{(Nq)}]$ (valid for $|g|<1$ in
the integration range), taking the 
$N\to \infty$ limit and resumming the resulting series. 
Noting now that $\Delta_N$ can be  written as $|\Delta_N|^2
={(|a|^2+|b|^2)}[1+r\sin(2Nq+\phi)]/[2 \sin^2{(q)}]$ 
 where $r\cos \phi={(ab^{\ast}+a^{\ast}b)}/{(|a|^2+|b|^2)},~r\sin
 \phi={(|b|^2-|a|^2)}/{(|a|^2+|b|^2)}$, we see that Eq.~(\ref{JSCR2})
 has the same structure as the left hand side of
 Eq.~(\ref{iden}). Hence using Eq.~(\ref{iden}) and after some
 simplification, we finally get:  
\bea
J_{C}&=&\f{\g k^2 k_B(T_L-T_R)}{\pi m}\int_{0}^\pi \f{\sin^2 q~ dq}{\Lambda -\Omega \cos q} \nn \\ 
&=&\f{\g k^2 k_B(T_L-T_R)}{m \Omega^2}(\Lambda-\sqrt{{\Lambda}^2-{\Omega}^2})~,\label{Class} \\
{\rm where}~~\Lambda&=& 2k(k-k')+{k'}^2+\f{(k_o+2k)\g^2}{m}~~{\rm
  and}~~\Omega= 2k(k-k') + \f{2k\g^2}{m}~.\nn
\eea
Eq.~(\ref{Class}) is the central result of this paper. 
We now show that  two different special cases lead to the RLL and N results.
First in the case of
fixed ends and without onsite potentials, {\emph{i.e.}} $k'=k$ and $k_o=0$,
we recover the RLL result \cite{Rieder67}:
\bea
J^{RLL}_{C}=\f{k k_B(T_L-T_R)}{2\g}\Big[1+\f{\nu}{2}-\f{\nu}{2}\sqrt{1+\f{4}{\nu}}~\Big]~~{\rm where}~~\nu=\f{mk}{\g^2}~.\label{RLL}
\eea
The case $k'=k, k_o \neq 0$ can also be obtained using the RLL
approach \cite{dharUNP} and agrees with the result in
Eq.~(\ref{Class}). 
In the other case of free ends, {\emph{i.e.}} $k'=0$, we get the
N result \cite{Nakazawa}: 
\bea
J^{N}_{C}=\f{k\g k_B(T_L-T_R)}{2(mk+\g^2)}\Big[1+\f{\lambda}{2}-\f{\lambda}{2}\sqrt{1+\f{4}{\lambda}}~\Big]~~
{\rm where}~~\lambda=\f{k_o \g^2}{k(mk+\g^2)}~.\label{NakO}
\eea

\section{Quantum mechanical case}
\label{sec:Quantum}
In the quantum case the heat current across a chain described by the
Hamiltonian Eq.~(\ref{ham}) and connected to Ohmic heat baths is given 
by \cite{DharRoy06}:
\bea
J_Q&=& \f{1}{\pi} \int_{-\infty}^\infty d \omega~ \hbar \omega
\cT_N (\omega) [f(\omega,T_L)-f(\omega,T_R)]~, \label{jqm}
\eea
where $f(\omega,T)=1/[e^{\hbar \omega/(k_B T)}-1]$ is the phonon
distribution function and $\cT_N$ is as given in Eq.~(\ref{JSCR}).
Here we consider the linear response regime where the applied
temperature difference $\Delta T = T_L-T_R << T$ with
$T=(T_L+T_R)/2$. Expanding the phonon distribution functions
$f(\omega,T_{L,R})$  about the mean temperature $T$ we get 
the following expression for the current:
\bea 
J_Q=\f{k_B(T_L-T_R)}{\pi}\int_{-\infty}^\infty d \omega~\left(\f{\hbar
  \om}{2 k_B T}\right)^2 {\rm{cosech}}^2 \Big(\f{\hbar \om}{2k_B
  T}\Big)~\mT(\om)~.\label{ocur1} 
\eea
We then proceed through the same asymptotic analysis as in the
previous section and get, in the
limit $N \to \infty$: 
\bea
J_Q&=&\f{\g k^2 \hbar^2 (T_L-T_R)}{4 \pi k_B m T^2}\int_{0}^\pi dq
\f{\sin^2 q}{\Lambda -\Omega \cos q}~ \om^2_{q}~{\rm{cosech}}^2
\Big(\f{\hbar \om_{q}}{2k_B T}\Big), \label{ocur2} \\ 
{\rm where}~~\om^2_{q}&=&[k_o+2k(1-\cos q)]/m~.\nn
\eea  
We are not able to perform the above integral exactly. Numerically it is easy
to obtain the integral for given parameter values and here we examine
the temperature dependence of the current (note that in the classical
case the current depends only on the temperature difference). In
Fig.(2) we plot the current as a 
function of temperature in three different cases (i) $k'=k, k_o=0$,
(ii) $k'=0, k_o=0$ and (iii) $k'=0, k_o \neq 0$. Particularly
interesting is the low temperature ($T << \hbar (k/m)^{1/2}/k_B$)
behaviour (shown in inset of 
Fig.(2)) which is very different  for the three cases.  The low
temperature behaviour can be obtained analytically by examining the
integrand  at small  $q$. We then find for the three different cases:
(i) $J_Q \sim T^3$, (ii) $J_Q \sim T$ and (iii) $J_Q \sim e^{-\hbar
  \omega_o/(k_B T)}/T^{1/2}$, where $\omega_o=(k_o/m)^{1/2}$.   
\begin{figure}[htb]
\vspace{1.5cm}
\includegraphics[width=13cm]{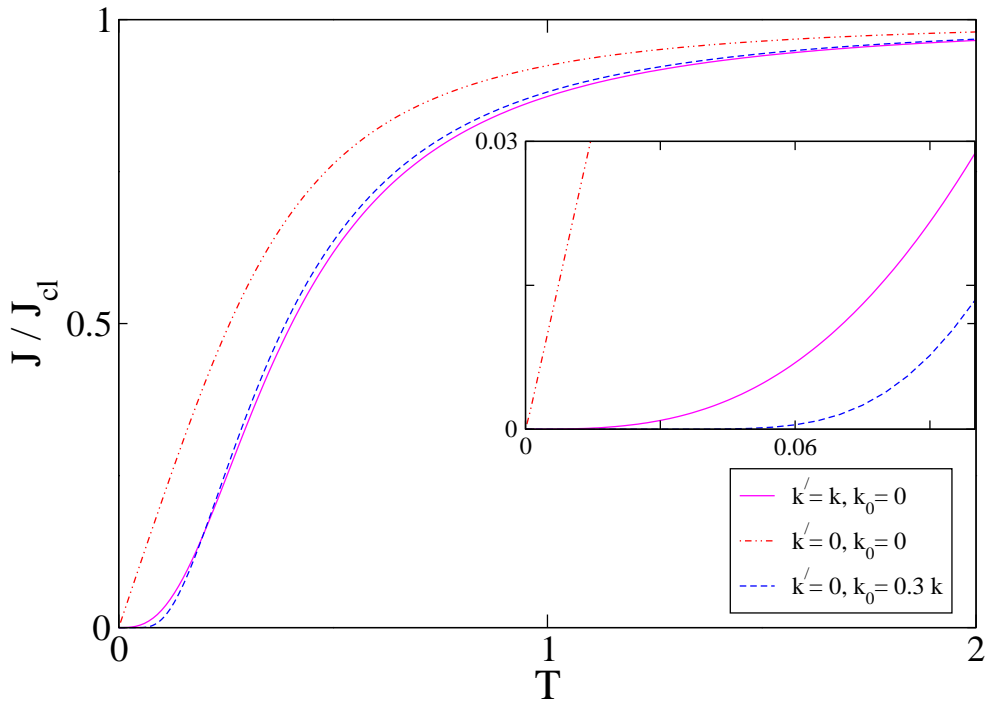}
\caption{Plot of the scaled heat current with temperature (in units of
 $\hbar (k/m)^{1/2}/k_B$) for three
 different parameter regimes (see text).  Inset shows the low
 temperature behaviour.}   
\label{fig2}
\end{figure}

\section{Higher dimensions}
\label{sec:Ddim}
Heat conduction in ordered harmonic lattices in more than one dimension was 
first considered by Nakazawa \cite{Nakazawa}. The problem can be
reduced to an effectively 
one-dimensional problem. For the sake of completeness we reproduce
their arguments 
here and also give the quantum generalization.

Let us consider a $d$-dimensional hypercubic lattice with lattice
sites labelled by the vector $\bl=\{l_\alpha\}, \alpha=1,2...d$, where
each $l_\alpha$ takes values from $1$ to $L_\alpha$. The total number of
lattice sites is thus $N=L_1L_2...L_d$. We assume that heat conduction
takes place in the $\alpha=d$ direction. Periodic boundary conditions
are imposed in the remaining $d-1$ transverse directions. The
Hamiltonian is described by a scalar displacement $X_\bl$ and as in
the $1D$ case we
consider  nearest neighbour harmonic interactions with a spring
constant $k$ and harmonic onsite pinning at all sites with spring
constant $k_o$. All boundary particles at $l_d=1$ and $l_d=L_d$ are
additionally pinned
by harmonic springs with stiffness $k'$ and follow Langevin dynamics
corresponding to baths at temperatures $T_L$ and $T_R$ respectively. 

Let us write $\bl=(\bl_t,l_d)$ where $\bl_t=(l_1,l_2...l_{d-1})$. Also
let ${\bf q}=(q_1,q_2...q_{d-1})$  with $q_\alpha= 2 \pi n/L_\alpha$
where $n$ goes from $1$ to $L_\alpha$. Then defining variables 
\bea
X_{l_d}({\bf q})=\f{1}{L_1^{1/2}L_2^{1/2}...L_{d-1}^{1/2}}\sum_{\bl_t}
X_{\bl_t,l_d} e^{i {\bf q}.\bl_t}~, 
\eea    
one finds that, for each fixed ${\bf q}$, $X_{l_d}({\bf q})$
($l_d=1,2...L_d$) satisfy  Langevin equations corresponding
to the $1$D Hamiltonian in Eq.~(\ref{ham}) with the  onsite spring
constant $k_o$ replaced by
\bea
\lambda({\bf q})=k_o+2[d-1-\sum_{\alpha=1,d-1} \cos{(q_\alpha)}]~.
\eea   
For $L_d\to \infty$, the heat current $J({\bf q})$ for each mode with
given ${\bf q}$ is 
then simply given by  Eq.(\ref{Class}) with $k_o$ replaced by
$\lambda_{\bf q}$.   In the quantum mechanical case we use
Eq.~(\ref{ocur2}).  The heat current per bond is then given by:
\bea
J=\f{1}{L_1L_2....L_{d-1}} \sum_{{\bf q}} J({\bf q})~.
\eea 
Note that the result holds for finite lengths in the transverse
direction. For infinite transverse lengths we get $J=\int...\int_0^{2 \pi} d
{\bf q} J({\bf q}) /(2 \pi)^{d-1}$~.

 \section{Summary}
\label{sec:concl}
In this paper we have derived the exact formula for the heat current
through an  ordered harmonic chain in the limit of infinite system
size. Our derivation is different from the methods used by RLL
\cite{Rieder67} and N \cite{Nakazawa}
and is for a slightly different version of the models studied by
them. We have presented the quantum mechanical generalization of
the results. In that case one gets, in the linear response regime, a
temperature dependent current with interesting low-temperature
behaviour. We have also stated the results for the general $d$-dimensional
case.


\begin{thebibliography}{10}
\bibitem{Rieder67} Z.~Rieder, J.~L. Lebowitz, and E.~Lieb,  
{\it Properties of a harmonic crystal in a stationary nonequilibrium
  state\/}, J. Math. Phys.~{\bf 8}, 1073 (1967).
\bibitem{Nakazawa} H. Nakazawa, {\it Energy Flow in Harmonic Linear
  Chain}, Progress of Theoretical Physics~ {\bf 39}, 236 (1968);
   {\it On the Lattice Thermal
  Conduction}, Progress of Theoretical Physics Supplement {\bf 45},
  231 (1970). 
\bibitem{dhar01} A. Dhar, {\it Heat conduction in the disordered
  harmonic chain revisited}, Phys. Rev. Lett. {\bf 86}, 5882 (2001).
\bibitem{DharRoy06} A. Dhar and D. Roy, {\it Heat transport in
  Harmonic lattices}, J.~Stat.~Phys.~{\bf 125}, 801 (2006). 
\bibitem{Casher71} A. Casher and J.L. Lebowitz, {\it Heat Flow in
  Regular and Disordered Harmonic Chains}, J. Math. Phys. {\bf 12},
  1701 (1971).
\bib{zurcher90} U. Zurcher and P. Talkner, {\it Quantum-mechanical
  harmonic chain attached to heat baths. II. Nonequilibrium
  properties}, Phys. Rev. A {\bf 42}, 3278 (1990).
\bibitem{saito00} K. Saito, S. Takesue, and S. Miyashita, {\it Energy
 transport in the integrable system in contact with various types of
 phonon reservoirs},  Phys. Rev. E {\bf 61}, 2397 (2000).
\bibitem{dhar03} A. Dhar and B. S. Shastry, {\it Quantum transport
  using the Ford-Kac-Mazur formalism}, Phys. Rev. B {\bf 67}, 195405 (2003).
\bibitem{segal03} D. Segal, A. Nitzan and P. Hanggi, {\it Thermal
  conductance through molecular wires}, J. Chem. Phys. {\bf
  119},  6840  (2003).
\bibitem{yamamoto06} T. Yamamoto and K. Watanabe, {\it Nonequilibrium
  Green's function approach to phonon transport in defective carbon
  nanotubes}, Phys. Rev. Lett. {\bf 96}, 255503 (2006).
\bibitem{wang06} J.-S. Wang, J. Wang and N. Zeng, {\it Nonequilibrium
  Green's function approach to mesoscopic thermal transport},  
Phys. Rev. B 74, 033408 (2006).
\bibitem{dharUNP} A. Dhar, unpublished notes.

\end{thebibliography}
\end{document}